\documentclass[11pt,twoside]{article}

\usepackage{asp2004}
\usepackage{psfig}
\usepackage{epsf}

\begin{document}

\title{Compact high-velocity clouds around the Galaxy and M31}

\author{Tobias Westmeier, Christian Br\"{u}ns, and J\"{u}rgen Kerp}

\affil{Radioastronomisches Institut der Universit\"{a}t Bonn, Auf dem H\"{u}gel~71, 53121~Bonn, Germany}

\begin{abstract}
  We have observed 11 compact high-velocity clouds in the 21-cm line of neutral hydrogen with the 100-m telescope in Effelsberg. Our observations show that most of the clouds have a rather complex morphology. Head-tail structures and bow-shock shapes suggest that many of the CHVCs are interacting with an ambient medium, providing additional evidence for a circumgalactic CHVC population. In order to verify the concept of a circumgalactic population directly, we have started an \ion{H}{i} blind survey to search for CHVCs in the neighbourhood of M31, using the Effelsberg telescope. We can confirm some of the extended \ion{H}{i} structures discovered by Thilker et al. (2004) with the Green Bank Telescope (GBT), whereas we have not yet found any cloud similar to the CHVCs observed near the Milky Way. Our non-detection suggests an upper limit for the distance of CHVCs from their host galaxies of about 60~kpc.
\end{abstract}

\section{Introduction}

High-velocity clouds (HVCs) were discovered by Muller, Oort, \& Raimond (1963) in the 21-cm line of neutral hydrogen. They are characterised by high radial velocities incompatible with a participation in the rotation of our Galaxy. HVCs can be found all across the sky in the form of extended complexes as well as compact, isolated clouds. One of the discussed hypotheses about their origin was introduced by Blitz et al. (1999) who suggested that HVCs might be the gaseous counterparts of the primordial dark-matter mini-haloes predicted by $\Lambda$CDM structure formation scenarios. This suggestion was seized by Braun \& Burton (1999) who identified 66 compact HVCs (CHVCs) in the Leiden/Dwingeloo Survey (LDS, Hartmann \& Burton 1997) with positional and kinematic characteristics being compatible with a distribution across the entire Local Group.

Recent observations and simulations have casted doubt on a Local Group population of HVCs. Zwaan (2000) investigated five galaxy groups with the Arecibo telescope but he could not detect any intra-group clouds. A similar survey by Pisano et al. (2004) with the Parkes telescope also had a negative result, indicating an upper limit of 160~kpc for the distance of HVCs from the Galaxy. In addition, Sternberg, McKee, \& Wolfire (2002) performed hydrodynamical simulations of CHVCs and found their circumgalactic model with typical distances from the Galaxy of the order of 150~kpc to be more consistent with the observed properties of the CHVCs. Similar results are obtained by Kravtsov, Gnedin, \& Klypin (2004).

We have studied 11~CHVCs in \ion{H}{i} with the Effelsberg telescope to search for indications of interaction with an ambient medium as additional evidence for CHVCs being located near the Galaxy. Furthermore, we have started an Effelsberg \ion{H}{i} survey of a large area around M31 to find a similar population of CHVCs around the second large spiral galaxy of the Local Group. Results from both projects are presented in this contribution.

\begin{table}[t!]
\caption{Physical parameters of the 11 observed CHVCs. $l$ and $b$ are the Galactic longitude and latitude of the column density maximum, $v_{\mathrm{LSR}}$ and $v_{\mathrm{GSR}}$ the column density weighted average radial velocities in LSR and GSR frames, $\Delta v$ the average line width (FWHM), $T_{\mathrm{B}}$ the observed peak brightness temperature, and $N_{\mathrm{HI}}$ the \ion{H}{i} peak column density. The last row gives the mean value $\langle x \rangle$ for each column. \label{tab_chvc}}
\begin{center}
{\small
\begin{tabular}{lrrrrr}
\hline
Name & $v_{\mathrm{LSR}}$ & $v_{\mathrm{GSR}}$ & $\Delta v$ & $T_{\mathrm{B}}$ & $N_{\mathrm{HI}}$ \\
CHVC $l \pm b$ & [$\mathrm{km \, s^{-1}}$] & [$\mathrm{km \, s^{-1}}$] & [$\mathrm{km \, s^{-1}}$] & [K] & [$10^{19} \; {\mathrm{cm}^{-2}}$] \\
\hline
CHVC 016.8$-$25.2    &     $-228$  &    $-171$  &     $14$  &      $3.1$  &      $5.6$ \\
CHVC 032.1$-$30.7    &     $-308$  &    $-207$  &     $30$  &      $1.3$  &      $6.0$ \\
CHVC 039.0$-$33.2    &     $-262$  &    $-147$  &     $22$  &      $1.6$  &      $8.0$ \\
CHVC 039.9$+$00.6    &     $-278$  &    $-137$  &     $32$  &      $0.7$  &      $4.5$ \\
CHVC 050.4$-$68.4    &     $-195$  &    $-133$  &     $27$  &      $1.3$  &      $4.7$ \\
CHVC 147.5$-$82.3    &     $-269$  &    $-254$  &     $22$  &      $2.2$  &      $8.0$ \\
CHVC 157.1$+$02.9    &     $-184$  &    $ -98$  &     $22$  &      $0.9$  &      $3.6$ \\
CHVC 172.1$-$59.6    &     $-235$  &    $-219$  &     $28$  &      $0.9$  &      $4.2$ \\
CHVC 218.1$+$29.0    &     $+145$  &    $ +27$  &     $ 6$  &      $2.8$  &      $3.2$ \\
CHVC 220.5$-$88.2    &     $-258$  &    $-263$  &     $22$  &      $1.0$  &      $3.7$ \\
CHVC 357.8$+$12.4    &     $-159$  &    $-167$  &     $27$  &      $1.5$  &      $6.4$ \\
\hline
$\langle x \rangle$  &     $-203$  &    $-161$  &     $23$  &      $1.6$  &      $5.3$ \\
\hline
\end{tabular}
}
\end{center}
\end{table}

\section{Effelsberg observations of compact high-velocity clouds}

We have observed 11~CHVCs in the 21-cm line emission of neutral hydrogen, using the 100-m radio telescope in Effelsberg. The clouds were selected on the basis of the CHVC catalogues published by Braun \& Burton (1999) and de~Heij, Braun, \& Burton (2002). We demanded an \ion{H}{i} column density ratio between previous, less sensitive Effelsberg observations of the clouds and the LDS data of $N_{\mathrm{HI}}^{\mathrm{Eff}} / N_{\mathrm{HI}}^{\mathrm{LDS}} \ge 3$. A high column density ratio implies the presence of compact sub-structures unresolved with the $36'$~HPBW of the Dwingeloo telescope but partly resolved with the significantly higher resolution of the Effelsberg telescope of $9'$~HPBW.

\subsection{Observations and data analysis \label{sect_obs}}

Each CHVC was mapped on a $9'$ grid ($4\farcm5$ in case of CHVC~218$+$29) with an integration time of 3~minutes for each position. The spectra were obtained in the inband frequency switching mode with a bandwidth of 6.3~MHz. We have 512~channels for each polarisation providing a velocity resolution of $2.6 \; \mathrm{km \, s}^{-1}$. The spectral baseline rms is about 50~mK, corresponding to a 1-sigma column density limit of about $2.4 \cdot 10^{17} \; \mathrm{cm}^{-2}$ per spectral channel.

In addition to the maps we took spectra along an appropriate axis of each cloud with a longer integration time of typically 10~minutes and a better angular sampling of $4\farcm5$ or $6\farcm4$. These deep cuts allow us to investigate the column density profile as well as the gradients in radial velocity and line width across each cloud in more detail. Along the deep cuts we reach a spectral baseline rms of about 30~mK which corresponds to a 1-sigma column density limit of about $1.4 \cdot 10^{17} \; \mathrm{cm}^{-2}$ per spectral channel.

\begin{figure}
  \plotone{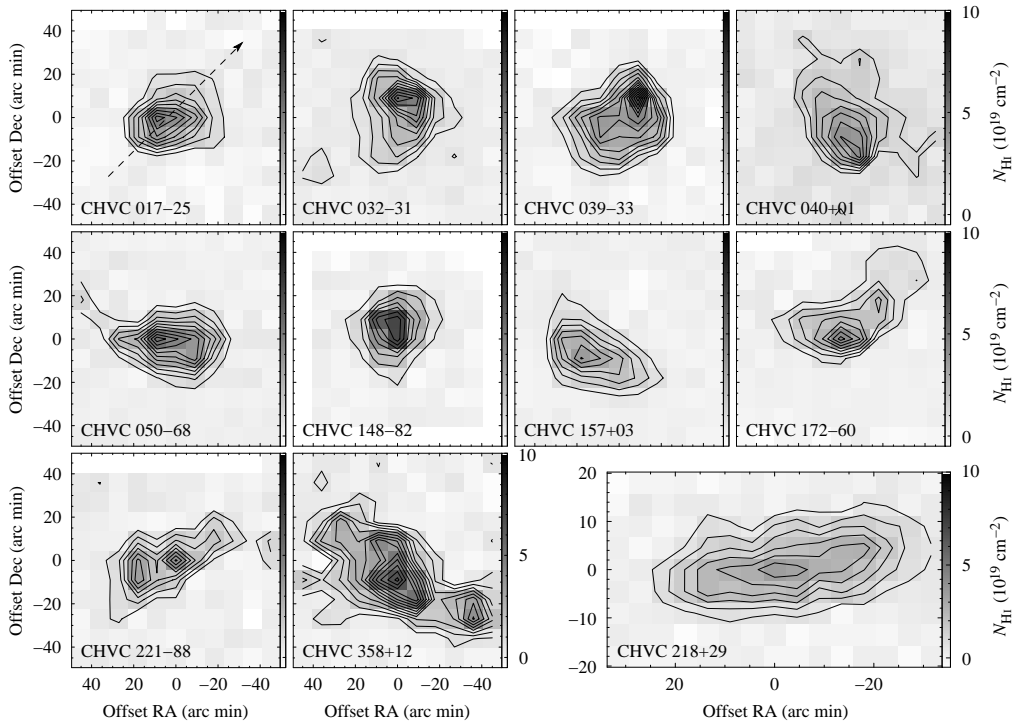}
  \caption{\ion{H}{i} column density maps of the 11 CHVCs observed with the Effelsberg telescope. Contours start at $5 \cdot 10^{18} \; \mathrm{cm}^{-2}$ ($1 \cdot 10^{19} \; \mathrm{cm}^{-2}$ in case of CHVC~040$+$01) with an increment of $5 \cdot 10^{18} \; \mathrm{cm}^{-2}$ ($1 \cdot 10^{19} \; \mathrm{cm}^{-2}$ in case of CHVC~148$-$82). \label{fig_chvcs_ov}}
\end{figure}

All spectra were calibrated using the S7 standard calibration source (Kal\-ber\-la, Mebold, \& Reif 1982). The spectra were corrected for stray radiation according to Kalberla, Mebold, \& Reich (1980). For most spectra a $3^{\mathrm{rd}}$ order polynomial was sufficient for baseline correction. To derive the physical parameters of the gas, we fitted a Gaussian to each spectral line and rejected all signals with a brightness temperature of $T_{\mathrm{B}} < 3 \, \sigma_{\mathrm{rms}}$. All radial velocities and line widths presented in this paper are based on these Gaussian fits.

\subsection{Results}

The physical parameters of the observed CHVCs are summarised in Table~\ref{tab_chvc}. The mean peak column density of about $5 \cdot 10^{19} \; \mathrm{cm}^{-2}$ is consistent with the value derived by Sternberg et al. (2002) from their hydrodynamical simulations of a circumgalactic CHVC population.

Most of the 11 CHVCs show a rather complex column density distribution instead of being spherically-symmetric (Fig.~\ref{fig_chvcs_ov}). Only one cloud, CHVC~148$-$82, has a spherically-symmetric appearance. Among the other clouds we find head-tail structures, bow-shock shapes, and also completely irregular morphologies without any radial or axial symmetry. These morphologies are suggestive of the CHVCs being distorted by external forces, e.g. by the ram pressure of an ambient medium through which the clouds are moving.

\begin{figure}
  \plotone{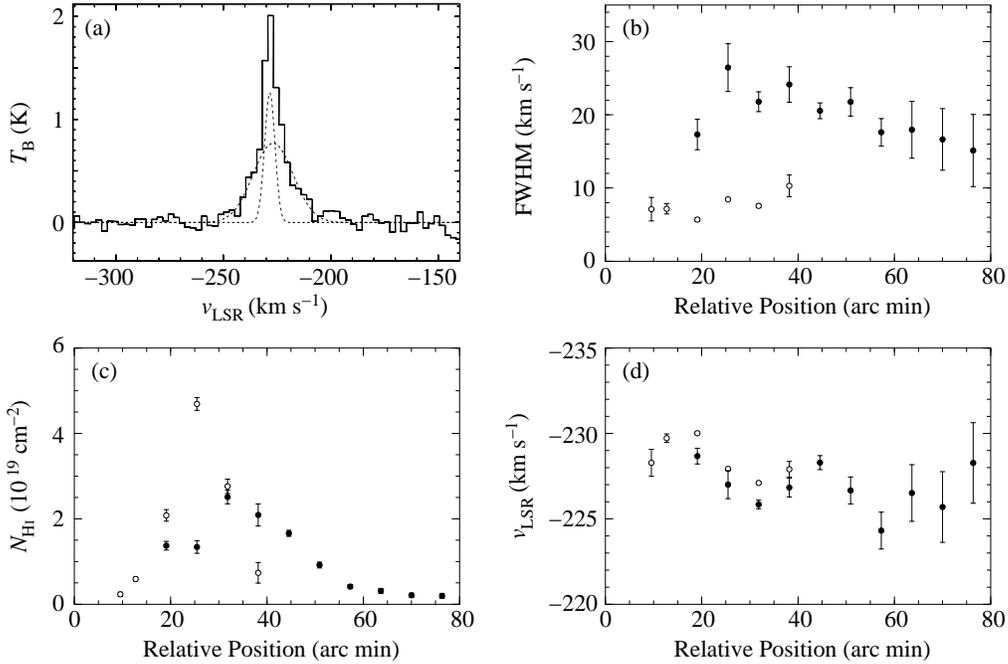}
  \caption{Results of a Gaussian decomposition of the spectra along the deep cut across CHVC~017$-$25. (a) Typical spectrum with a broad and a narrow Gaussian component (dashed curves). (b) Line widths, (c) \ion{H}{i} column densities, and (d) radial velocities of the cold ($\circ$) and the warm ($\bullet$) gas component along the cut. \label{fig_twogauss}}
\end{figure}

Additional evidence for the possibility of ram-pressure interaction between some of our CHVCs and an ambient medium comes from the physical properties of the gas observed along the deep cuts. One example, CHVC~017$-$25, shall be discussed here in more detail. In Fig.~\ref{fig_twogauss}~(a) we plotted a typical spectrum along the deep cut indicated by the arrow in the map in Fig.~\ref{fig_chvcs_ov}. The line profile resembles the superposition of a narrow and a broad Gaussian component, indicating the presence of a cold and a warm neutral gas. We have made the attempt to fit, where possible, two Gaussians to the spectral lines along the cut to investigate the parameters of the cold and the warm gas component separately. The results are shown in Fig.~\ref{fig_twogauss}~(b) to (d).

Fig.~\ref{fig_twogauss}~(b) shows the observed line widths along the cut for the cold gas (open circles) and the warm gas (filled circles). Both gas components are well separated with line widths of $6 \ldots 10 \; \mathrm{km \, s}^{-1}$ FWHM for the cold neutral medium and $16 \ldots 26 \; \mathrm{km \, s}^{-1}$ FWHM for the warm neutral medium. These values correspond to upper limits for the kinetic temperature of $800 \ldots 2200 \; \mathrm{K}$ and $5600 \ldots 15000 \; \mathrm{K}$, respectively. In Fig.~\ref{fig_twogauss}~(c) we plotted the \ion{H}{i} column density profiles for both gas components along the cut. The cold component is quite compact while the warm gas is much more extended. Both seem to be spatially separated with the warm gas forming an extended, faint tail in one direction. This result suggests that the envelope of CHVC~017$-$25 is being stripped off the core by the ram pressure of a surrounding medium. The radial velocities of the two gas components are displayed in Fig.~\ref{fig_twogauss}~(d). The small velocity difference between the cold and the warm gas, however, is not significant considering the velocity resolution of our data.

Our observations suggest that some of the 11~CHVCs are indeed distorted by the ram-pressure forces of an ambient medium through which the clouds are moving. Hydrodynamical simulations of interacting HVCs by Vieser (2001) and Quilis \& Moore (2001) have shown that densities of the ambient medium of the order of $n \approx 10^{-4} \; \mathrm{cm}^{-3}$ are required to account for head-tail structures like the ones found in our observations. Thus, the detected signs of possible ram-pressure interactions provide additional evidence for a CHVC population in the neighbourhood of our Galaxy. 

\section{Effelsberg survey of M31}

\begin{figure}
  \plotone{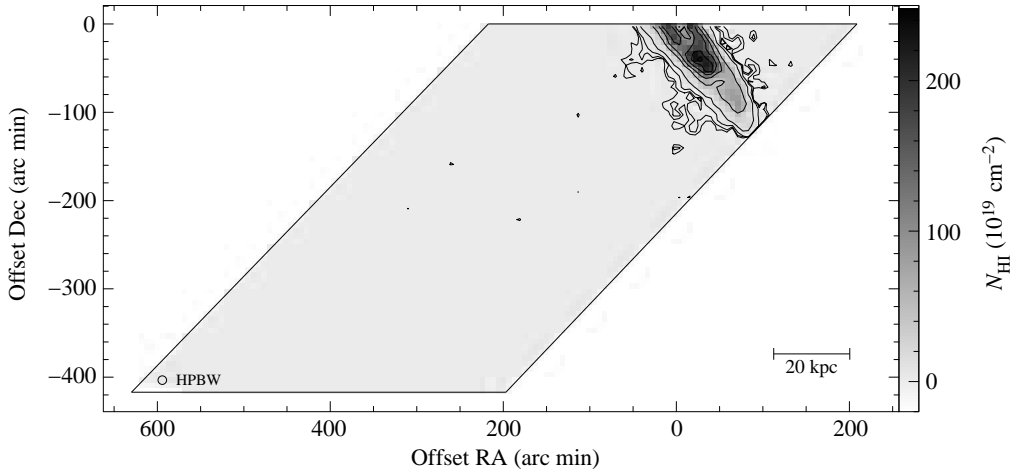}
  \caption{\ion{H}{i} column density map of the area around M31 mapped with the Effelsberg telescope, so far (shaded area). Contours are drawn at 0.75, 1.25, 2.5, 10, 50, 100, 150, 200, and $250 \cdot 10^{19} \; \mathrm{cm}^{-2}$. \label{fig_m31_ov}}
\end{figure}

In order to directly verify the concept of a circumgalactic population of CHVCs as the gaseous counterparts of primordial dark-matter mini-haloes, we have started an \ion{H}{i} blind survey in direction of M31 using the Effelsberg telescope. The aim of this survey is to search for CHVCs in the vicinity of M31. The area mapped so far extends about $10^{\circ}$ away from the centre of M31 in south-eastern direction, corresponding to a projected distance of 130~kpc (see Fig.~\ref{fig_m31_ov}). Thus, our survey is complementary to the GBT survey of Thilker et al. (2004) which extends to about $50 \; \mathrm{kpc}$ in projection from the centre of M31.

\subsection{Observations and data analysis}

The observations were carried out with the same specifications as described in Sect.~\ref{sect_obs} except that we used the normal frequency switching mode. Again, we obtain a spectral baseline rms of about 50~mK at the original velocity resolution of $2.6 \; \mathrm{km \, s}^{-1}$. The corresponding 1-sigma column density limit is $2.4 \cdot 10^{17} \; \mathrm{cm}^{-2}$ per spectral channel. Adjacent spectra are spatially separated by $9'$ matching the HPBW of the Effelsberg telescope.

Some of the \ion{H}{i} features found near the disk of M31 were re-observed with higher sensitivity and partly with higher angular sampling to study the properties of the gas in more detail. For these observations we made use of the new 8192-channel autocorrelator and the inband frequency switching method with a bandwidth of 10~MHz. We obtain a velocity resolution of $0.5 \; \mathrm{km \, s}^{-1}$ for both polarisations with 4096 spectral channels each. The spectral baseline rms in these re-observed maps is about 50~mK, corresponding to a very low 1-sigma column density limit of about $5 \cdot 10^{16} \; \mathrm{cm}^{-2}$ per spectral channel.

\subsection{Results}

\subsubsection{Confirmation of the large HVCs}

\begin{figure}
  \plotone{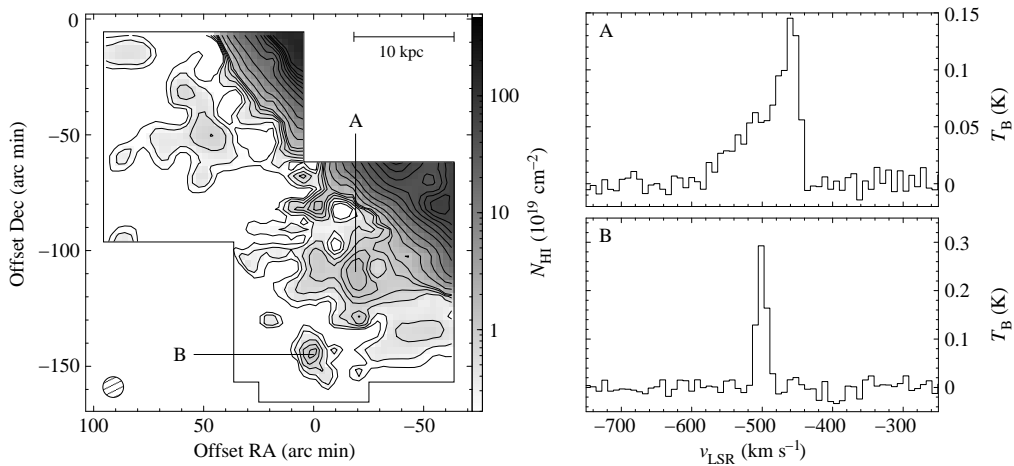}
  \caption{\ion{H}{i} column density map of the clouds found near the eastern and south-eastern edge of the disk of M31. The centre of M31 is located at offset $(0,0)$. Contours are drawn at 0.1, 0.25, 0.5, 0.75, 1, 1.25, 2.5, 5, 10, 20, 30, 40, 50, 60, 70, 80, and $100 \cdot 10^{19} \; \mathrm{cm}^{-2}$. Two example spectra from the positions labelled with A and B are shown on the right. The complex line profile at position~A indicates the presence of compact sub-structure unresolved with the $9'$~HPBW of the Effelsberg telescope. The cloud at position~B shows the highest intensities and quite narrow lines of only about $15 \; \mathrm{km \, s}^{-1}$ FWHM. \label{fig_fields}}
\end{figure}

Some of the large \ion{H}{i} structures discovered by Thilker et al. (2004) with the GBT were covered by our map and could be confirmed. These large-scale structures located south-east of the disk of M31 were re-observed with higher sensitivity and partly with a better angular sampling to study their properties in more detail. A combined column density map and example spectra are shown in Fig.~\ref{fig_fields}. The \ion{H}{i} emission found east of the centre of M31 (upper-left part of the map in Fig.~\ref{fig_fields}) is quite diffuse and faint with a peak column density of about $7 \cdot 10^{18} \; \mathrm{cm}^{-2}$. The radial velocities of the gas of $v_{\mathrm{LSR}} \approx -350 \; \mathrm{km \, s}^{-1}$ are slightly more negative than the systemic velocity of M31. The emission detected south-east of M31 (lower-right part of the map in Fig.~\ref{fig_fields}) is much more pronounced with a peak column density of nearly $1.5 \cdot 10^{19} \; \mathrm{cm}^{-2}$. The emission appears quite smooth but many spectra show very complex, non-Gaussian profiles (e.g. Fig.~\ref{fig_fields}~A), indicating the presence of compact sub-structures which are unresolved with the $9'$ HPBW of the Effelsberg telescope. In many cases, neighbouring spectra have completely different line shapes and radial velocities, suggesting that several small clumps with different radial velocities exist in this direction.

A remarkable object is the cloud labelled with B in the map in Fig.~\ref{fig_fields}. It is isolated and with about $10'$~FWHM barely resolved with the Effelsberg telescope. The total \ion{H}{i} mass of the cloud is $(5.1 \pm 0.2) \cdot 10^5 \; M_{\odot}$ if we assume a distance of 780~kpc (Stanek \& Garnavich 1998). The \ion{H}{i} lines are remarkably narrow for our 2-kpc beam with only about $15 \; \mathrm{km \, s}^{-1}$ FWHM, corresponding to an upper limit for the kinetic temperature of about 5000~K. The profiles of the spectral lines are not perfectly Gaussian but show signs of a narrow component indicating the presence of a cold core embedded in an envelope of warm gas.

\subsubsection{The search for CHVCs}

Apart from the large-scale structures found in projection close to the disk of M31 we discovered about 30 signals all across the field being compact and isolated in position and velocity. Most of these CHVC candidates were very faint with peak intensities of only a few $\sigma_{\mathrm{rms}}$. Therefore, it was necessary to re-observe the positions of all CHVC candidates with the Effelsberg telescope to confirm their existence. With these re-observations, which were carried out in early 2004, none of the CHVC candidates could be confirmed. The most likely explanation for our failure to confirm the candidates is that the detected signals had been caused by radio frequency interference.

This negative result has an important impact on the concept of circumgalactic CHVCs as the gaseous counterparts of primordial dark-matter mini-haloes. One possible interpretation of our result is that M31 does not host a population of CHVCs at all or that at least such a population has properties different from the Galactic population. If, on the other hand, we assume that an M31 CHVC population exists and that it has properties similar to those of the population observed around the Milky Way we can use our non-detection to estimate upper limits for the distances of the CHVCs from their host galaxies as well as their \ion{H}{i} masses and sizes. The distance of the Milky Way CHVCs from the observer can be estimated by
\begin{equation}
  d^{\mathrm{MW}} = \frac{d^{\mathrm{M31}} \tan \left( \sqrt{\varphi_{\mathrm{Eff}}^2 T_{\mathrm{B}}^{\mathrm{M31}} / T_{\mathrm{B}}^{\mathrm{MW}}} \right)}{\tan \varphi^{\mathrm{MW}}} \label{eqn_1}
\end{equation}
where the indices MW and M31 denote the corresponding parameters of the Milky Way CHVCs and the M31 CHVCs, respectively. Furthermore, $d$ is the distance of the CHVCs from the observer, $T_{\mathrm{B}}$ is the observed brightness temperature, $\varphi$ is the angular size of the CHVCs, and $\varphi_{\mathrm{Eff}}$ denotes the HPBW of the telescope. In addition, the assumption is made that the M31 CHVCs are unresolved with the Effelsberg telescope. From our observations of Galactic CHVCs we get average values of $T_{\mathrm{B}}^{\mathrm{MW}} \! \approx 0.75 \; \mathrm{K}$ over $\varphi^{\mathrm{MW}} \! \approx 30'$. Moreover, $\varphi_{\mathrm{Eff}} = 9'$ and $d^{\mathrm{M31}} = 780 \; \mathrm{kpc}$ (Stanek \& Garnavich 1998). If we consider the \ion{H}{i} lines of the CHVCs to be resolved with a velocity resolution of $20.8 \; \mathrm{km \, s}^{-1}$ we obtain $T_{\mathrm{B}}^{\mathrm{M31}} < 50 \; \mathrm{mK}$ from our non-detection of M31 CHVCs. This assumes that we are able to detect and confirm a $3 \, \sigma_{\mathrm{rms}}$ signal with a more sensitive re-observation of the same position. Inserting the above values into Eqn.~\ref{eqn_1} yields an upper limit for the distance of the Milky Way CHVCs from the Sun of
\begin{equation}
  d^{\mathrm{MW}} \! < 60 \; \mathrm{kpc} \, . \label{eqn_2}
\end{equation}
This result suggests that CHVCs might lie much closer to their host galaxies than previously thought. It is in agreement with the latest H$\alpha$ detections towards a number of Galactic CHVCs by Tufte et al. (2002) and Putman et al. (2003), suggesting distances of some 10~kpc. We note, however, that our survey is to some extent incomplete because of beam-by-beam sampling and the limited overall coverage. From the distance limit in Eqn.~\ref{eqn_2} we can derive upper limits for the sizes $D$ and the \ion{H}{i} masses $M_{\mathrm{HI}}$ of CHVCs:
\begin{equation}
  D < 0.5 \; \mathrm{kpc} \, , \quad M_{\mathrm{HI}} < 6 \cdot 10^4 \, M_{\odot} \, ,
\end{equation}
again assuming a $3 \, \sigma_{\mathrm{rms}}$ detection limit and a line width of $23 \; \mathrm{km \, s}^{-1}$ FWHM which is the mean value derived from our observations of Galactic CHVCs.

\acknowledgments{T. Westmeier acknowledges support by the Deutsche Forschungsgemeinschaft (DFG) through project number KE~757/4-1. Based on observations with the 100-m telescope of the MPIfR (Max-Planck-Institut f\"{u}r Radioastronomie) at Effelsberg.}

\end{document}